# Large-scale $\alpha^2$-dynamo in low-mass stars and brown dwarfs

Gilles Chabrier[1,2,*] and Manfred Küker[2]

[1] Ecole Normale Supérieure de Lyon, CRAL (UMR CNRS 5574), 69364 Lyon Cedex 07, France
    chabrier@ens-lyon.fr
[2] Astrophysikalisches Institut Potsdam, 14482 Potsdam, Germany
    mkueker@aip.de

**Abstract.** We develop a model based on three dimensional mean-field magnetohydrodynamics computations for the generation of large scale magnetic fields in fully convective objects like low-mass stars, brown dwarfs and possibly gaseous planets. The dynamo process is of $\alpha^2$ type and thus differs from the shell-dynamo at work in more massive stars. The $\alpha^2$ dynamo is found to become supercritical for Coriolis numbers $\Omega^\star \gtrsim 1$, i.e. Rossby numbers $\mathcal{R}o \lesssim 10$. It generates a large-scale, non-axisymmetric, steady field that is symmetric with respect to the equatorial plane. Saturation of the $\alpha^2$-generated field at the equipartition field strength yields strengths of several kilo-Gauss, in agreement with observations of active M dwarfs, and provides a qualitative explanation for the observed activity saturation in late M stars. For brown dwarfs with a conductive core, as occurs at the center of the most massive and oldest of these objects, we have also studied an $\alpha^2\Omega$ dynamo, i.e. the effect of differential rotation. In this case the field is predominantly toroidal, axisymmetric, and oscillatory, like the solar magnetic field.

The topology of the field in the fully convective objects exhibits a high order multipole character that differs from the aligned dipole field generated by the $\alpha\Omega$ dynamo. The strong reduction of the dipolar component due to the non-axisymmetry of the field should considerably reduce the Alfven radius and thus the efficiency of magnetic braking, providing an appealing explanation for the decreasing angular momentum loss rate observed in low-mass stars and brown dwarfs. This may have also implications for cataclysmic variables below the period gap. In spite of this large-scale field, the decreasing conductivity in the dominantly neutral atmosphere of these objects may prevent the current generation necessary to support a chromosphere and thus activity. An observational signature of the present model would be (i) asymmetry of the chromospheric activity, contrary to the spatially uniform activity expected from small-scale turbulent dynamo and (ii) the absence of cycles in uniformly rotating (fully convective) low-mass objects.

**Key words:** Magnetohydrodynamics - Turbulence - Stars: magnetic fields - Stars: activity - Stars: rotation - Stars: chromosphere - Stars: low mass, brown dwarfs - Stars: cataclysmic variables

[*] Johann Wempe prize

## 1. Introduction

Various observations have revealed ongoing chromospheric activity among M-type stars, with strong $H_\alpha$ emission. The $H_\alpha$ equivalent width, a primary indicator of stellar activity for these objects, increases significantly for spectral types $\gtrsim$ M2-M3 and then exhibits a rather uniform distribution with a mean level $\log(L_{H_\alpha}/L_{bol}) \sim -4.0$ (Gizis et al. 2002, Mohanty & Basri 2003). Most, if not all M dwarfs with spectral-type $> M4$ ($T_{eff} \lesssim 3100$ K, $M_{bol} \gtrsim 11$) show significant chromospheric activity around the aforementioned value. On the other hand, X-ray surveys of the solar neighborhood and of several open clusters have identified numerous M dwarfs as persistent faint X-ray sources, with typical luminosities of the order of or larger than the average solar luminosity, $L_X \sim 10^{27}$ erg s$^{-1}$, i.e. $L_X/L_{bol} \sim 10^{-3}$, up to a spectral type of M8 (Randich 2000 and references therein). This includes stellar objects and young ($\lesssim 10^7$-$10^8$ yr) brown dwarfs (Neuhäuser et al. 1999, Tsuboi et al. 2003, Briggs et al. 2004). This suggests that late M dwarfs are as efficient coronal emitters as other cool stars in terms of $L_X/L_{bol}$, which expresses the level of X-ray activity. On the other hand, there is observational evidence for a decrease in the chromospheric activity by at least one order of magnitude with respect to late M dwarfs, that begins at spectral types $\gtrsim$ M9, i.e. $T_{eff} \sim 2200$ K, $M_{bol} \gtrsim 13.5$ and continues to L dwarfs, in spite of the fact that these objects are very rapid rotators, with $v \sin i > 10$ km s$^{-1}$ (Delfosse et al. 1998, Reid et al. 2002, Mohanty & Basri 2003). This has led to the suggestion that chromospheric activity drops at some point with decreasing mass and/or temperature (Stauffer et al. 1994, Gizis et al. 2000). This is supported by the fact that brown dwarfs have been observed to exhibit strong $H_\alpha$ emission in very young associations, with ages $\lesssim 10$ Myr, whereas none of these objects has been shown to show activity in older clusters, with ages $\gtrsim 50$ Myrs. Similarly, no late L or T spectral type object has been found to show persistent activity (Mohanty & Basri 2003). On the other hand, flares showing strong $H_\alpha$ emission have been observed in L and T brown dwarfs, with small duty cycles of order 1% (Liebert et al. 2003 and references therein), suggesting that the absence of continuous coronal or chromospheric activity does not preclude the presence of a strong magnetic field.

These observations, in particular the changes in the magnetic activity around spectral types $\sim$ M2 and $\sim$ M9 provide important information about the magnetic activity in low-mass stars and substellar objects. First, the spectral type domain M2-M3 encompasses the mass range where M dwarfs are expected to become fully convective, $m \lesssim 0.35\,M_\odot$ for solar abundance (Chabrier & Baraffe 1997). Therefore, magnetic field generation does not disappear in fully convective objects like very-low-mass stars and brown dwarfs. This raises the question of a large scale magnetic field generation in fully convective objects and a change in the dynamo mechanism around $0.35\,M_\odot$, different to the rotation-driven dynamo believed to be operative in the Sun and early M-dwarfs with a radiative core. Second, as mentioned above, the decreasing magnetic activity for late-M, L and T dwarfs but the presence of strong flares confirms the generation of a large scale magnetic field in the interior of these objects but suggests a decreasing *coupling* between the field and atmospheric motions for objects cooler than $T_{eff} \sim 2200$ K. This latter issue has been addressed by Mohanty et al. (2002), following an earlier suggestion by Meyer & Meyer-Hofmeister (Meyer & Meyer-Hofmeister 1999). These authors suggest that the drop-off in the very late-M and L dwarfs stems from the strongly decreasing electrical conductivity in these cool, predominantly neutral atmospheres, beginning with late M dwarfs.

Another crucial piece of information is provided by the observed rotation velocities of these objects. It is now well established that the usual monotonic rotation-activity connection observed for solar-type and early M ($\lesssim$ M3) stars breaks down for cooler objects. While high chromospheric and coronal activity is associated with high rotation velocities for the former ones, M9 and later objects exhibit low activity levels in spite of rapid rotation (Delfosse et al. 1998, Gizis et al. 2002, Mohanty & Basri 2003). The $L_X/L_{bol}$ and $L_{H_\alpha}/L_{bol}$ ratios both correlate with $v\sin i$ for $v\sin i \lesssim$ 4-5 km s$^{-1}$, corresponding to a rotation period $P \sim 3$ days, and then saturate at a limit $\log(L_X/L_{bol}) \simeq -3$ and $\log(L_{H_\alpha}/L_{bol}) \sim -4.0$, respectively, above this rotational velocity threshold. This implies that the intrinsic coronal and chromospheric emission of low-mass stars and brown dwarfs does not increase with increasing rotation above the threshold limit. This suggests a saturation relation in terms of the Rossby number, i.e. the ratio of the rotational period, $P = 2\pi R/v_{rot} = 2\pi/\Omega$, over the convective turnover time, $\tau_{\rm conv}$, $\mathcal{R}o = \frac{P}{\tau_{\rm conv}}$ (see §2). Note that the high rotation velocities observed for most M dwarfs and essentially all L dwarfs is usually interpreted as a spin-down timescale that increases with decreasing mass. Since angular momentum loss is generally associated with magnetic breaking, this phenomenon is rooted in the same fundamental problem of magnetic field generation and dissipation in these objects.

These data suggest the following conclusions : (i) dynamo action in fully convective objects is as efficient as in stars with a radiative core, (ii) whatever this dynamo mechanism, it is bounded by the saturation condition for late-type stars, (iii) flares or large-scale features like spotted areas definitely make a large-scale field necessary, (iv) not only does activity not increase with increasing rotation above the threshold limit but M9 and later objects display low activity levels in spite of rapid rotation. It is the aim of this paper to examine these issues, all rooted in the fundamental question: what very active dynamo mechanism is responsible for the large scale magnetic field generation in cool, fully convective objects such as very-low-mass stars, brown dwarfs and possibly giant planets ?

## 2. The dynamo

For a fully convective star, a fossil field can survive only over a timescale $\tau_d \approx R^2/\eta \sim$ 10-100 years, where $\eta \equiv \eta_t$ is the turbulent magnetic diffusivity and $R$ the stellar radius. A dynamo process is thus necessary to generate and amplify the magnetic field. It is widely accepted that magnetic activity in the Sun and possibly in early M stars results from the generation of a large-scale toroidal field by the action of differential rotation on a (weak) poloidal field at the interface between the convective envelope and the radiative core, where differential rotation is strongest, known as the tachocline (Parker 1975, Spiegel & Weiss 1980, Spiegel & Zahn 1992). This region has also been suggested as the seat of the dynamo because magnetic flux tubes are stable against buoyancy for much larger field strengths than in the convection zone (Parker 1975). In this case the field generation by shear in the tachocline dominates that of the helicity in the convection zone and the magnetic field is predominantly toroidal. This is the so-called $\alpha\Omega$ dynamo-generation, which predicts a strong correlation between activity and rotation, or more exactly between the field generation rate and the Rossby number $\mathcal{R}o$, as observed for F, G and early M stars. The release of the magnetic stresses, created through the dragging of field lines by fluid motions in the upper atmosphere, provides energetic support for a corona and a chromosphere and generates activity. Efficiency of $\alpha\Omega$ dynamo correlates strongly with decreasing Rossby number, so that large rotational velocities lead to high activity levels. A pure interface dynamo has, however, the problem of producing too many toroidal field belts and too short cycle periods (Rüdiger & Brandenburg 1995). A dynamo operating at the bottom of the convection zone also has the shortcoming of producing the wrong correlation between the radial and azimuthal field components (Stix 1976). Both these problems are solved by the advection-dominated dynamo, which is essentially an $\alpha\Omega$ dynamo, but with shear and $\alpha$ effect operating in different layers, which are connected by the meridional flow (Choudhuri et al. 1995, Küker et al 2001). The cycle period is then determined by the flow speed rather than the dynamo number.

While the absence of a radiative core, i.e. of a region of weak buoyancy and strong differential rotation, precludes in principle the generation of a large-scale magnetic field by solar-type dynamo, an $\alpha^2\Omega$ mechanism can not be ruled out.

An alternative process for fully convective stars, namely a dynamo generated by a turbulent velocity field that would generate chaotic magnetic fields in the absence of rotation, has been proposed by Durney et al. (Durney et al. 1993). The turbulent velocity field can generate a self-maintained *small-scale* magnetic field providing the magnetic Reynolds number $Re = (vl/\eta)^2$ is large enough. The length and time scales of this field are comparable to that of turbulence, i.e. mixing length $l$ and convective turnover time $l/v$, and thus yields only small-scale fields. Rotation is not essential in this case but it increases the generation rate of the field, although mildly. Moreover, although such a small-scale magnetic field can maintain a certain level of magnetic activity, it can not produce the high levels of activity observed in M stars and brown dwarfs, as discussed in §1. One might thus expect a sharp decrease of the rotation-activity connection around ∼ M3, where full convection sets in, contrary to what is observed. We thus explore the possibility of generating a *large-scale* magnetic field by a pure $\alpha^2$ dynamo process.

### 2.1. Mean-field magnetohydrodynamics

In the present calculations, we treat the global magnetic field within the framework of mean-field magnetohydrodynamics. This approach is particularly suited to situations in which the magnetic field exhibits complex structures in space and time, such as when the fluid motions are of turbulent nature. In the mean-field approach, the magnetic as well as the velocity fields are split into an averaged and a fluctuating part:

$$\boldsymbol{B} = \bar{\boldsymbol{B}} + \boldsymbol{B}', \qquad \boldsymbol{u} = \bar{\boldsymbol{u}} + \boldsymbol{u}' \quad (1)$$

where, in the present context, $\boldsymbol{u}'$ is the fluctuating velocity of turbulence. In mean field MHD, the mean field is identified as the large-scale field. The induction equation for the mean magnetic field then reads:

$$\frac{\partial \bar{\boldsymbol{B}}}{\partial t} = \nabla \times \left( \bar{\boldsymbol{u}} \times \bar{\boldsymbol{B}} + \overline{(\boldsymbol{u}' \times \boldsymbol{B}')} \right). \quad (2)$$

The second term on the right-hand side of Eq.(2) is the turbulence-generated electromotive force. It it usually assumed to be a functional of the mean magnetic field $\bar{\boldsymbol{B}}$ that can be expanded in a Taylor series (Krause & Rädler 1980):

$$\overline{(\boldsymbol{u}' \times \boldsymbol{B}')}_i = \alpha_{ij} \bar{B}_j + \beta_{ijk} \frac{\partial \bar{B}_j}{\partial x_k} + \dots \quad (3)$$

In case of a slowly rotating star, the tensors $\alpha_{ij}$ and $\beta_{ijk}$ become isotropic and the induction equation (2) takes the well-known form:

$$\frac{\partial \bar{\boldsymbol{B}}}{\partial t} = \nabla \times (\bar{\boldsymbol{u}} \times \bar{\boldsymbol{B}} + \alpha \bar{\boldsymbol{B}} - \eta \nabla \times \bar{\boldsymbol{B}}) \quad (4)$$

The first term on the right-hand side of Eq.(4) describes induction by the mean gas flow, including the generation of the toroidal field in differentially rotating objects ("$\Omega$ effect"). The second term denotes the amplification of the mean magnetic field by helical turbulence ("$\alpha$ effect"). The third term is purely diffusive, with $\eta$ denoting the (turbulent) magnetic diffusivity coefficient. Note that while the diffusivity term exists even in isotropic turbulence, the $\alpha$-term exists only if the turbulence medium is stratified *and* rotates, i.e. only in the case of inhomogeneous, non-isotropic turbulence. Rotation is thus important to generate the $\alpha$-effect, first by generating kinetic and thus magnetic helicity, and second because of its effect on turbulence. So even an $\alpha^2$-dynamo depends on the rotation rate, or more exactly on the Rossby number.

### 2.2. The electromotive force for rapidly rotating stellar objects

Eq. (4) is valid for slow rotation, i.e. if the convective turnover time $\tau_{\rm conv}$ is small compared to the rotation period $P$ of the star. As mentioned previously, the ratio of these time scales is measured by the Rossby number $\mathcal{R}o$ or its inverse the Coriolis number $\Omega^* = 4\pi/\mathcal{R}o = 2\tau_{\rm conv}\Omega$, which is large in the case of rapid rotation. Based on the second order correlation approximation (SOCA, also known as first order smoothing approximation), the coefficients $\alpha_{ij}$ and $\beta_{ijk}$ can be derived from some general assumptions about the underlying turbulent velocity field and the stellar stratification (Rüdiger & Hollerbach 2004). They finally become functions of the Coriolis number.

We distinguish two types of stratification that are important for the electromotive force in different regions of the star, namely the stratifications of density and turbulence intensity. They are represented by the vectors:

$$\boldsymbol{G} = \nabla \log \rho \quad (5)$$

and

$$\boldsymbol{U} = \nabla u_t, \qquad u_t = \sqrt{\boldsymbol{u}'^2}. \quad (6)$$

In a spherically symmetric object both vectors point in the radial direction but the orientation may differ. While $\boldsymbol{G}$ always points outwards, $\boldsymbol{U}$ points out of the convection zone, i.e. *inwards* at the lower boundary of a superficial convection zone. In the case of the Sun, it points inwards throughout the whole convection zone because the convection velocity increases with radius.

The presence of a large-scale magnetic field will affect the motion of the gas and hence the turbulent electromotive force (EMF). So far there is no theory of the turbulent EMF in a rapidly rotating, strongly magnetized stratified fluid. As we are primarily interested in the question of whether a large-scale field can be generated at all, we adopt the expressions valid for the case of weak fields. The back-reaction of the field is then taken into account by a simple $\alpha$ quenching prescription (see below).

#### 2.2.1. The $\alpha$ effect

The $\alpha$ effect is the field generation by the mean helicity of the gas motion, which is the consequence of the combined effects of rotation and stratification. Both the stratification of density and of the turbulence contribute to the effect. While being an important ingredient in theories of an $\alpha\Omega$ dynamo in the solar overshoot layer (e.g. Rüdiger & Brandenburg 1995), the $\alpha$ effect due to the stratification of the turbulence is of minor importance in the bulk of the solar convection zone, or in a fully convective star (Krivodubskij & Schultz 1993). The contribution of the density stratification reads (Kitchatinov & Rüdiger

1992, Rüdiger & Hollerbach 2004):
$$\alpha^\rho_{ij} = -\delta_{ij}(\boldsymbol{G\Omega})\alpha^\rho_1 - (G_i\Omega_j + G_j\Omega_i)\alpha^\rho_2$$
$$-(G_i\Omega_j - G_j\Omega_i)\alpha^\rho_3 - \frac{\Omega_i\Omega_j}{\Omega^2}(\boldsymbol{G\Omega})\alpha^\rho_4 \qquad (7)$$

### 2.2.2. Turbulent buoyancy

Besides the field-generating $\alpha$ effect the density stratification causes a pure transport effect that under certain circumstances might affect the dynamo. It is the turbulent buoyancy (Kitchatinov 1991),
$$\boldsymbol{\mathcal{E}}_{\text{buoy}} = \boldsymbol{u}_{\text{mag}} \times \langle\boldsymbol{B}\rangle - \frac{2}{\Omega^2}(\boldsymbol{u}_{\text{mag}} \times \boldsymbol{\Omega})(\boldsymbol{\Omega} \cdot \langle\boldsymbol{B}\rangle) \qquad (8)$$
where
$$\boldsymbol{u}_{\text{mag}} = -\boldsymbol{G}_\perp \tau_{\text{conv}} u_t^2 \psi_\phi(\Omega^*) \qquad (9)$$
with
$$\boldsymbol{G}_\perp = \boldsymbol{G} - \boldsymbol{\Omega}(\boldsymbol{G}\cdot\boldsymbol{\Omega})/\Omega^2, \qquad (10)$$
and
$$\psi_\phi = \frac{1}{4\Omega^{*2}}\left[\frac{\Omega^{*2}+3}{\Omega^*}\arctan\Omega^* - 3\right] \qquad (11)$$
It is convenient to include the turbulent buoyancy in the $\alpha$ effect, wich in cylindrical coordinates $(s,\phi,z)$ then reads:
$$\alpha = \begin{pmatrix} \hat{\alpha}\psi\cos\theta & 0 & \hat{\alpha}\psi_\varpi\sin\theta \\ 0 & \hat{\alpha}\psi\cos\theta & u_m \\ \hat{\alpha}\psi_\varpi\sin\theta & u_m & \hat{\alpha}\psi_z\cos\theta \end{pmatrix} \qquad (12)$$
where
$$u_m = \sqrt{\boldsymbol{u}_{\text{mag}}^2}. \qquad (13)$$
Note that $u_m$ contains a factor $\sin\theta$ and is thus symmetric with respect to the equatorial plane. Like $\psi_\phi$, $\psi$, $\psi_z$, and $\psi_\varpi$ are functions of the Coriolis number:
$$\psi = \frac{1}{\Omega^{*3}}\left(\Omega^{*2}+6 - \frac{6+3\Omega^{*2}-\Omega^*}{\Omega^*}\arctan\Omega^*\right) \qquad (14)$$
$$\psi_z = \frac{1}{\Omega^{*3}}\left(-\frac{10\Omega^{*2}+12}{1+\Omega^{*2}} + \frac{2\Omega^{*2}+12}{\Omega^*}\arctan\Omega^*\right) \qquad (15)$$
$$\psi_\varpi = \frac{2}{\Omega^{*3}}\left(3 - \frac{\Omega^{*2}+3}{\Omega^*}\arctan\Omega^*\right) \qquad (16)$$

$\alpha$ is isotropic in the case of slow rotation but becomes strongly anisotropic for fast rotation:
$$\alpha^\rho \approx c_\alpha G \tau_{\text{conv}} u_t^2 \frac{\pi}{2}\cos\theta \begin{pmatrix} 1 & 0 & 0 \\ 0 & 1 & 0 \\ 0 & 0 & 0 \end{pmatrix} \qquad (17)$$
for $\Omega^* \gg 1$. Note that no field component parallel to the axis of rotation ($z$ axis) will be generated by Eq.(17).

We restrict ourselves to a purely kinematic dynamo model, due to the helicity of the fluid motion, without nonlinear effects like an $\alpha$ effect caused by magnetic helicity or the twisting of flux tubes (Ferriz-Mas et al. 1994). This approach is valid for field strengths below the equipartition value,
$$B_{\text{eq}} = \sqrt{4\pi\rho u_t^2}. \qquad (18)$$
The growth of the magnetic field strength is limited by introducing a simple $\alpha$ quenching prescription,
$$\alpha_{ij}(B) = \frac{\alpha_{ij}(0)}{\sqrt{1+B/B_{\text{eq}}^2}}, \qquad (19)$$

as back reaction of the magnetic field on the fluid motions. Eq.(19) stops the growth of the field when equipartition with the turbulent motions is reached.

### 2.2.3. Magnetic diffusivity

The magnetic diffusivity tensor for a rotating fluid reads (Kitchatinov et al. 1994, Rüdiger & Hollerbach 2004):
$$\eta_{ijk} = \eta_0(\phi_1+\phi_2)\epsilon_{ijk} + \eta_0(\phi_1-\phi_2)\epsilon_{ijl}\frac{\Omega_l\Omega_k}{\Omega^2}. \qquad (20)$$
where $\eta_0$ is the magnetic diffusivity for a nonrotating fluid and the coefficients $\phi_i$ are dimensionless functions of the Coriolis number. In the limit of slow rotation, $\phi_1 = 1$ and $\phi_2 = 0$ and $\eta$ becomes isotropic. For rapid rotation, the diffusivity decreases with the Coriolis number as $1/\Omega^*$. The magnitude of the magnetic diffusivity is given by the common mixing-length relation
$$\eta_0 = c_\eta l u_t, \qquad c_\eta \approx 1/3. \qquad (21)$$
$$\phi_1 = \frac{3}{4\Omega^*}\left(-1+\frac{\Omega^{*2}+1}{\Omega^*}\arctan\Omega^*\right) \qquad (22)$$
$$\phi_2 = \frac{3}{2\Omega^*}\left(1-\frac{\arctan\Omega^*}{\Omega^*}\right) \qquad (23)$$

### 2.3. Numerics

The EMF for rapidly rotating convection zones becomes strongly anisotropic with the rotation axis as the preferred direction. In case of the $\alpha$ effect this means that its component parallel to the axis of rotation vanishes. On the other hand, differential rotation must be expected to be weak or completely absent. The dynamo is therefore of $\alpha^2$ type, which in case of anisotropic $\alpha$ effect is very likely to produce non-axisymmetric fields. We thus have to carry out the computations in three dimensions. We employ the code used in Küker & Rüdiger (1999) which is based on an explicit finite-difference scheme after Elstner et al. (1990). As the code works in cylindrical polar coordinates, the star is embedded in a cylinder filled with a poorly conducting passive medium to avoid boundary conditions on the stellar surface. The value of the magnetic diffusivity coefficient in the surrounding cylinder slightly depends on the Coriolis number but is larger than inside the star by about a factor of 10. This configuration mimics a star surrounded by vacuum. The boundary conditions imposed on the surfaces of the cylindrical container then must ensure that the Poynting flux through these surfaces vanishes and the magnetic field is divergence-free. These conditions are achieved by requiring:
$$B_s = B_\phi = \frac{\partial B_z}{\partial z} = 0 \qquad (24)$$
on the top and bottom surfaces while
$$\frac{\partial}{\partial s}(sB_s) = B_\phi = B_z = 0 \qquad (25)$$
on the outer surface. We choose a radius of two stellar radii and a total height of four stellar radii for the container to avoid a strong influence of the boundaries on the stellar interior. As usual, we start the computation with a small, divergence-free seed field with mixed parity that contains an axisymmetric as well as a non-axisymmetric component. The evolution of the field is then followed until either the total field energy has decreased by several orders of magnitude and still increases exponentially with time, or a state is reached where the field is either stationary or oscillatory with constant amplitude. All computations are done with 121 × 61 × 9 grid points in the $z$, $s$, and $\phi$ coordinates. The code and the choice of the seed field allow solutions of any parity or symmetry. In order to examine how sensitive the results are to the setup conditions of the numerical calculations, we have repeated some of the computations with a larger simulation box (doubling both the heigth and the radius) and/or with an increased resolution.

| No. | age /Gyr | $T_{eff}$ (K) | radius / $10^{10}$ cm | $\eta_0$ / ($10^{10}$cm$^2$ s$^{-1}$) | $G$ | $B_{\rm eq}$ / kG |
|---|---|---|---|---|---|---|
| I | 0.032 | 2800 | 1.48 | 35 | 3 | 4.0 |
| II | 0.65 | 2000 | 0.68 | 3.7 | 3, 5 | 4.0 |
| III | 10 | 940 | 0.54 | 0.46 | 5, 10 | 1.0 |

**Table 1.** Basic parameters for the brown dwarf models

*2.4. Stratification*

In this paper, we use as a template model for our investigations a 0.06 M$_\odot$ brown dwarf model, of which inner structure is representative of fully convective low-mass stars and brown dwarfs (Chabrier & Baraffe 1997, 2000). The characteristic conditions are given in Table 1. In the following, we refer to these conditions as characteristic of models I, II and III, distinguished by different ages of 0.032 Gyr, 0.65 Gyr, and 10 Gyr, respectively. Figure 1 shows the variation with radius of the quantities that determine the turbulent EMF, namely the density, the equipartition field strength, the convective turnover time and the convection velocity of the brown dwarf at three different ages. As the model uses the local mixing-length approximation, all quantities are strongly stratified. While the inverse density scale height and the convection velocity increase with increasing radius, the convective turnover time decreases very rapidly and the equipartition field strength is roughly constant except close to the center and the surface of the star. The flat inner region for the 10 Gyr model illustrates the presence of a conductive core in very cool and massive brown dwarfs, where conduction eventually takes over convection to transport energy (Chabrier et al. 2000).

Running the calculations with the stratifications as plotted in Fig. 1 would require a very high spatial resolution in both the $s$ and $z$ directions and thus make the computations unreasonably expensive. We therefore simplify the model by working with constant values for all quantities (including the magnitude of the density stratification vector $G$ which of course must remain finite). The dashed lines in Fig. 1 indicate the values used in the computations, as listed in Table 1. For Models II and III we made additional runs with $G = 5$, which in case of Model II is more representative of the outer shell than the value $G = 3$ taken at half the radius. For Model III a value $G = 5$ refers to a radius close to the *base* of the convection zone rather than at its middle. The value $G = 10$ shown in the figure represents a radius halfway between the bottom and the top of the convection zone. We then carry out a series of runs with increasing rotation rates. So far the rotation pattern of brown dwarfs is unknown. Recent observational studies (Barnes et al 2005) as well as theoretical models (Küker & Rüdiger 2005) suggest a decrease of the surface shear with spectral type i.e. with increasing convection zone depth. We therefore assume rigid rotation for the fully convective objects. As long as no differential rotation is included, the stellar rotation rate only enters via the Coriolis number, which is always a large number, because of the long convective turnover times. As symmetry requires the $\alpha$ effect to vanish at the stellar center we set it to zero for radii smaller than 10 percent of the stellar radius. *For Model III, we also carry out computations with differential rotation*. To keep the model as simple as possible we use a simple radial shear term:

$$\Omega_s = \Omega_0(0.9 + 0.1\max(x, 1.0)), \qquad (26)$$

where $\Omega_0 = 10^{-6}$s$^{-1}$ and $x$ is the fractional stellar radius. The rotation rate thus increases linearly with the radius inside the star and is constant outside (corotation of the outer medium with the surface of the star avoids an artificial shear layer).

**3. Results**

As mentioned above, we keep all parameters constant except the rotation rate, which is varied as the input parameter. The turbulent velocities for the three brown dwarf model stratifications are taken from the average values of the convective velocities displayed in Fig. 1. The corresponding convective turnover times are $\tau_{\rm conv} = 100$ d for models I and III and $\tau_{\rm conv} = 160$ d for model II. For a Coriolis number $\Omega^\star \sim 1$, these values correspond to rotation periods $P \sim 1260$ days and 2010 days, repectively.

Larger convective time scales or smaller rotation periods yield larger Coriolis numbers and thus even more supercritical situations. When there is no differential rotation, the rotation rate only enters via the Coriolis number. Starting with slow rotation and increasing the rotation rate, the dynamo begins to work once the critical value for the Coriolis number is reached. We then carry out a series of computations to study how the field structure changes as the Coriolis number is further increased, for each of the three models. The results are summarized in Table 2. In this table, S and A modes are defined by the symmetry of the field with respect to the equatorial plane with for S modes: $B_s(z) = B_s(-z), B_\phi(z) = B_\phi(-z), B_z(z) = -B_z(-z)$ and for A modes: $B_s(z) = -B_s(-z), B_\phi(z) =$ $-B_\phi(-z), B_z(z) = B_z(-z)$. As for the azimuthal symmetry field, axisymmetric solutions obey the condition $(\frac{\partial}{\partial \phi} = 0)$ for all field components and thus correspond to an azimuthal quantum number m=0.

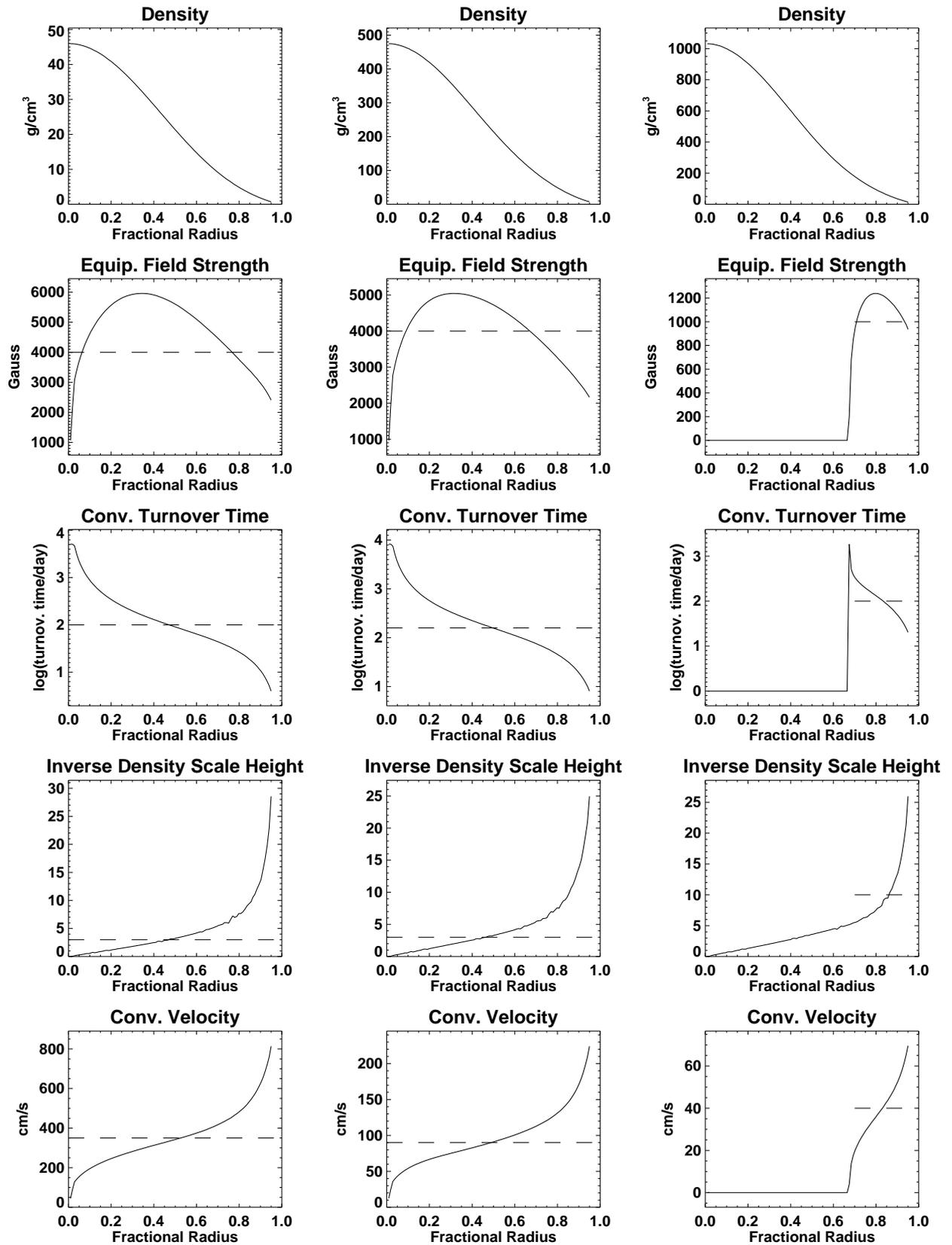

**Fig. 1.** Stratifications of a 0.06 $M_\odot$ brown dwarf at ages of 0.032 Gyr (left column), 0.65 Gyr (center), and 10 Gyr (right). The diagrams show the variation of the density, the equipartition field strength, the convective turnover time, the inverse density scale height, and the convection velocity, as functions of the fractional stellar radius. Note that the actual radius varies with the age of the star (Chabrier & Baraffe 2000). The dashed lines indicate the default values used in the computations, $G = 3$ for Model I (left) and Model II (center) and $G = 10$ for Model III (right).

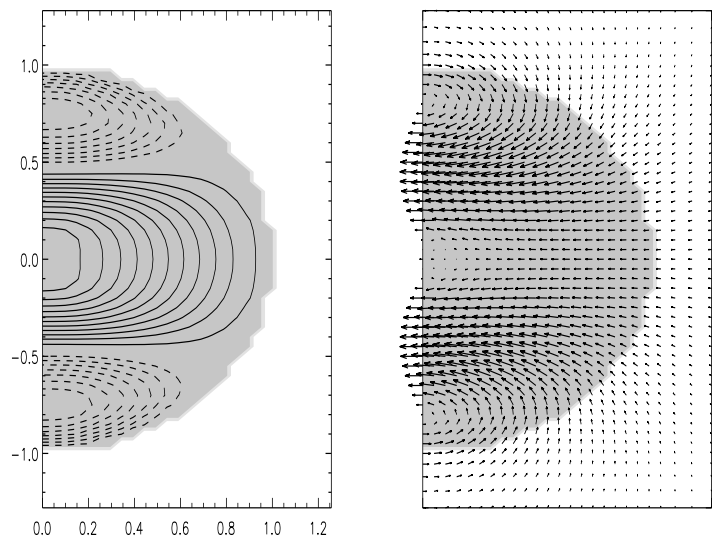

**Fig. 2.** Magnetic field in the $\phi = 0$ half plane for Model I and a Coriolis number $\Omega^* = 2$. The grey-shaded area marks the volume of the star. Solid contours denote a field component pointing into the paper (positive $\phi$ component) while dashed lines indicate that the field points out of the paper. This solution is not axisymmetric, hence the plot only represents one particular half plane.

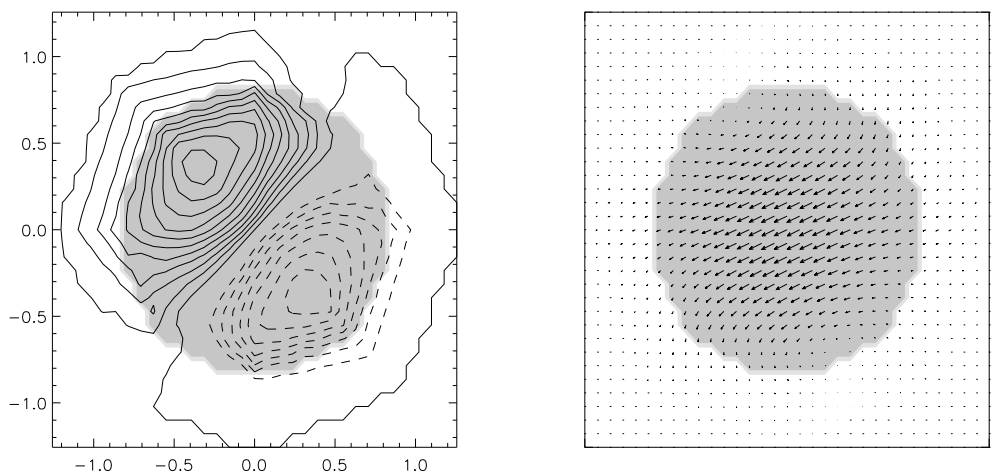

**Fig. 3.** Same model as in Fig.2, but now in a plane of constant positive $z$. Note that in this representation solid contours mark areas where the field points out of the paper (positive $z$ component). Due to the symmetry of the field, the z component vanishes in the $z = 0$ plane. Had a value of the same modulus but opposite sign been chosen for z, the right picture would be the same while in the left part the line styles would be interchanged.

For the youngest object (Model I), the critical Coriolis number for dynamo action is about $\Omega^* = 1.3$, i.e. a Rossby number $\mathcal{R}o = 9.7$. With a convective turnover time of 100 days, this corresponds to an angular velocity $\Omega \sim 7.5 \times 10^{-7}\mathrm{s}^{-1}$ or a rotation period of 967 days. Above this rotation rate, we find a steady field with S1 symmetry, i.e. symmetric with respect to the equatorial plane and *no axisymmetric contribution*. The field geometry for $\Omega^* = 2$ is shown in Figs. 2 and 3. Figure 2 shows a cut through the $\phi = 0$ half plane and Fig. 3 a cut through the northern hemisphere at constant value of the $z$ coordinate. The projection into this plane as well as the normal component of the field show a structure resembling a tilted dipole. Note, however, that the field is *not axisymmetric* with respect to either the axis of rotation (z axis) or with respect to any other axis but shows mirror-symmetry with respect to the equatorial plane (S1 mode). We have not carried out a multipole expansion of the field but the resulting symmetry implies that *it consists only of quadrupolar or higher order terms*.

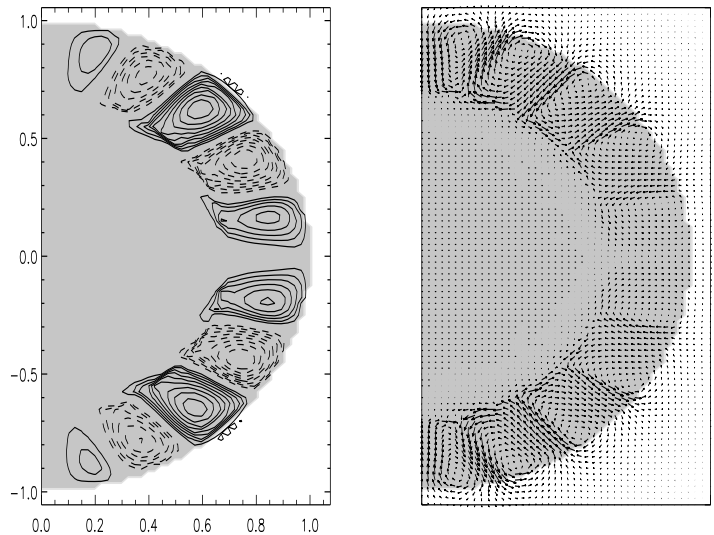

**Fig. 4.** Magnetic field for Model III with $G = 5$, a Coriolis number $\Omega^* = 5$, and differential rotation. The representation is the same as in Fig.2. In the right picture, the arrows mark poloidal field. The length of the vectors is proportional to the strength of the (poloidal) field.

The critical Coriolis number for the onset of the dynamo in Model II lies between 1 and 1.3 for $G = 3$ and between 0.7 and 0.8 for $G = 5$. For all Coriolis numbers above the critical value we find the same S1 symmetry and field strengths of the same order of magnitude as in model I. The $G = 5$ model has a stronger $\alpha$ effect and therefore generates a stronger magnetic field at the same Coriolis number. As the $\alpha$ effect is determined by the density stratification, the results for the $G = 3$ are the same as for Model I.

The oldest object (model III) differs from the two younger ones in the fact that the core is degenerate enough to become conductive instead of convective. For $G = 5$ and no differential rotation the onset of the dynamo is now found between $\Omega^* = 1$ and $\Omega^* = 1.5$. The solution has S1 symmetry, as the previous fully convective cases of models I and II. Adding differential rotation lowers the critical Coriolis number to a value below $\Omega^* = 0.5$ i.e. $\mathcal{R}o \gtrsim 25$. The field is *now* axisymmetric and oscillatory, like the solar magnetic field, but symmetric (S0 instead of A0 symmetry). We recall that in a mean-field approach, the Cowling theorem does not apply, and one can have an axisymetric mean field *and* generate dynamo. At $\Omega^* = 10$, we find a solution with mixed parity and for $\Omega^* = 20$ the solution has A0 geometry. For the larger value $G = 10$ and no differential rotation dynamo action sets in between $\Omega^* = 0.6$ and $\Omega^* = 0.8$. The geometry is S1 and the field is steady. For the highly supercritical case $\Omega^* = 20$ the parity switches to dipole. With differential rotation the geometry changes from S1 to S0 (i.e. the field becomes axisymmetric) for the moderately supercritical case and the solution is oscillatory. Figure 4 shows a snapshot of the field geometry for $G = 5$ and $\Omega^* = 5$.

For $\Omega^* = 20$ the solution is of the A0 type now (axisymmetric dipole).

The generation of magnetic fields of S1 symmetry by $\alpha^2$ dynamos in spherical geometry has been found before. Rädler et al. (1990) studied linear $\alpha^2$ dynamos with both isotropic and anisotropic $\alpha$ effect. They found the A0 mode to be the first to become supercritical for isotropic $\alpha$ effect, but in case of an anisotropic $\alpha$ effect the S1 mode has the lowest critical Reynolds number.

Rüdiger et al. (2003) studied $\alpha^2$ dynamos with $\alpha$ tensors of various forms, including the limiting case for very rapid rotation of the form we use (their $\hat{\alpha}_z = 0$, $\lambda = 0$ case). They found that the solution for fully convective stars is always completely non-axisymmetric, i.e. of S1 or A1 symmetry, unless the $\alpha$ effect is highly anisotropic, strongly concentrated to the equatorial region, and restricted to a thin outer shell. While the isotropic $\alpha$ effect favours solutions of the S1 type for deep convection zones, the critical dynamo number is the same for A1 and S1 solutions in the case of a shallow convection zone and mixed modes can occur (Moss et al. 1990).

For an anisotropic $\alpha$ effect with a $\cos\theta$ dependence on latitude and a deep convection zone, the critical dynamo numbers of the A1 and S1 modes are very similar (11.8 and 11.7, respectively) and the S1 mode is therefore only weakly preferred. For a thin convection zone, the critical dynamo numbers of the A1 and S1 mode are the same, as for isotropic $\alpha$. While a thin spherical shell surrounding a perfectly conducting core (which the dynamo-generated field cannot penetrate) favours axisymmetric solutions, in our case we find only non-axisymmetric solutions for the $\alpha^2$ dynamo. This is in agreement with Rüdiger et al. (2003), who found the same for an $\alpha \propto \cos\theta$. The situation changes when differential rotation is added. In this case axisymmetric solutions are found. Another property of thin shell dynamos is that the field geometry contains high-order multipole components, especially in case of axisymmetric solutions. Typically, there is a sequence of belts of alternating polarity, with the width in the horizontal directions approximately the same as the depth of the field-generating layer. We do not search for the critical dynamo numbers and we have not studied pure A0 and S0 modes separately. Instead we start with a mixed geometry and evolve the field until a stationary (or oscillatory) state is reached. We find the S0 mode to be preferred, but in some cases a small A0 component persists and in one case the field geometry remains in a state of mixed symmetry. This suggests that the critical dynamo numbers are almost identical. This is not suprising as the field geometries of the A0 and S0 modes do not differ much in the case of many field belts (multipole of high order), as in Fig. 4.

We made a number of separate runs with *double resolution* either in the $R-z$ plane or in the azimuthal direction, which all confirmed the results obtained with lower resolution. For runs with a *larger simulation box*, two gave different results to the original runs. For Model I, $\Omega^\star = 2$, the solution switched from S1 to A1 and for Model III, $G = 5$, $\Omega^\star = 1$, we found A0 instead of S0. We attribute this to the fact that the critical dynamo numbers for the A and S modes of the same azimuthal symmetry are very similar. *We did not find an axisymmetric solution instead of a non-axisymmetric one or vice versa in any case*. All results are summarized in Table 2.

## 4. Discussion and conclusion

We have explored the possibility of generating a *large-scale* field by a pure $\alpha^2$-effect in the interior of fully convective objects like very-low-mass stars, brown dwarfs or gaseous planets. In the $\alpha^2$-dynamo, helicity is generated by the action of the Coriolis force on the convective motions in a rotating, stratified fluid. Although for rapid rotation ($\Omega^\star \gg 1$), the $\alpha^2$-term does not depend explicitly on the Coriolis number, the magnetic diffusivity does, with $\eta \propto 1/\Omega^\star$ (see §2.2.3). Therefore, the $\alpha^2$-effect depends on the Coriolis number, or the equivalent Rossby number $\mathcal{R}o = 4\pi/\Omega^\star$. We have carried out numerical MHD simulations in the framework of mean-field magnetohydrodynamics. We find that the $\alpha^2$-dynamo is clearly supercritical for $\Omega^\star \gtrsim 1$, $\mathcal{R}o \lesssim 10$. It generates a large-scale, non-axisymmetric, steady (co-rotating) field that is symmetric with respect to the equatorial plane. These calculations thus show that *$\alpha^2$-dynamo can efficiently generate a large-scale magnetic field in the interior of fully convective objects*. The growth of the field is limited by $\alpha$ quenching and thus the dynamo produces field strengths of the order of the equipartition field strength, i.e. several kiloGauss, in agreement with observations of active M dwarfs. In this model, the final field geometry is essentially always the same, i.e. exhibits a high order multipole character. The field strength, on the other hand, depends on the stellar parameters and rotation rate: larger Coriolis numbers, i.e. smaller Rossby numbers, yield stronger fields.

The observed continuous transition in rotation and activity at the fully convective boundary suggests that the $\alpha^2$-dynamo

| Model | $\Omega^\star$ | Parity | Symmetry | time dep. | Field energy |
|---|---|---|---|---|---|
| I | 1.2 | | 1 | decay | |
| | 1.3 | S/S | 1 | steady | $4.1 \times 10^6$ |
| $G=3$ | 1.5 | S/S | 1 | steady | $1.4 \times 10^7$ |
| | 2 | S/A | 1 | steady | $4.1 \times 10^7$ |
| | 5 | S/S | 1 | steady | $2.0 \times 10^8$ |
| | 10 | S | 1 | steady | $4.5 \times 10^8$ |
| | 20 | S | 1 | steady | $9.3 \times 10^8$ |
| | 50 | S | 1 | steady | $2.3 \times 10^9$ |
| II | 1.0 | | | decay | |
| | 1.3 | S | 1 | steady | $4.1 \times 10^6$ |
| $G=3$ | 1.5 | S | 1 | steady | $1.4 \times 10^7$ |
| | 2 | S | 1 | steady | $4.1 \times 10^7$ |
| | 5 | S | 1 | steady | $2.0 \times 10^8$ |
| | 10 | S | 1 | steady | $4.5 \times 10^8$ |
| | 20 | S | 1 | steady | $2.3 \times 10^9$ |
| II | 0.7 | | | decay | |
| | 0.8 | S/S | 1 | steady | $1.5 \times 10^6$ |
| | 1 | S/S | 1 | steady | $1.8 \times 10^7$ |
| $G=5$ | 2 | S/S | 1 | steady | $1.2 \times 10^8$ |
| | 5 | S/S | 1 | steady | $4.1 \times 10^8$ |
| | 10 | S | 1 | steady | $8.2 \times 10^8$ |
| | 20 | S | 1 | steady | $1.6 \times 10^9$ |
| | 50 | S | 1 | steady | $3.9 \times 10^9$ |
| III | 1 | | | decay | |
| | 1.5 | S | 1 | steady | $2.9 \times 10^5$ |
| $G=5$ | 2 | S | 1 | steady | $1.1 \times 10^6$ |
| | 5 | S | 1 | steady | $7.1 \times 10^6$ |
| | 10 | S | 1 | steady | $1.7 \times 10^7$ |
| | 20 | S | 1 | steady | $3.6 \times 10^7$ |
| III | 0.4 | | | decay | |
| | 0.5 | S | 0 | oscil | $2.4 \times 10^5$ |
| | 1 | S/A | 0 | oscil | $3 \times 10^6$ |
| $G=5$ | 2 | S/S | 0 | oscil | $1 \times 10^7$ |
| DR | 5 | S/S | 0 | oscil | $3 \times 10^7$ |
| | 10 | M | 0 | oscil | $8 \times 10^7$ |
| | 20 | A | 0 | oscil | $7.5 \times 10^8$ |
| III | 0.5 | | | decay | |
| | 0.7 | S | 1 | steady | $9.7 \times 10^4$ |
| $G=10$ | 1 | S | 1 | steady | $1.1 \times 10^6$ |
| | 2 | S | 1 | steady | $5.9 \times 10^6$ |
| | 5 | S | 1 | steady | $1.9 \times 10^7$ |
| | 10 | S | 1 | steady | $3.9 \times 10^7$ |
| | 20 | A | 1 | steady | $7 \times 10^7$ |
| III | 0.6 | | | decay | |
| | 0.7 | S | 0 | oscil | $3 \times 10^5$ |
| $G=10$ | 1 | S | 0 | oscil | $2.5 \times 10^6$ |
| DR | 2 | S | 0 | oscil | $1.3 \times 10^7$ |
| | 5 | S | 0 | oscil | $4.5 \times 10^7$ |
| | 10 | S | 0 | oscil | $1.3 \times 10^8$ |
| | 20 | A | 0 | oscil | $3 \times 10^8$ |

**Table 2.** Field geometry and energy for different ages and rotation rates. In the parity field, S (A) means symmetric (antisymmetric) with respect to the equatorial plane and M denotes solutions without a definite symmetry. In the cases where two parity values are given, the second one refers to the model with the outer boundaries farther away from the star. The numbers 0 and 1 in the azimuthal Symmetry field denote purely axisymmetric solutions and complete absence of an axisymmetric component, respectively. The term 'DR" indicates the series of models include differential rotation (Eq.(26)). The energies are given in arbitrary units that differ between the models

is already at work in the convection zones of the more massive stars[1]. Eventually it becomes strong enough near mid M spectral types to yield saturation between rotation and activity even for slow rotation rates. Indeed, as shown by Rüdiger & Kitchatinov (1993), a turbulent $\alpha^2$ dynamo exhibits the same basic properties as the $\alpha\Omega$ dynamo, namely the magnetic flux increases with rotation, because of the effect of rotation on turbulence, and saturates at high rotation rates, when the $\alpha^2$ mode becomes supercritical. This saturation effect, however, cannot be quantified in the present mean field approach which estimates the $\alpha$-quenching energy as the equipartition energy. A quantitative determination requires non-linear (dissipative) processes, not included in the present calculations. The present mean field calculations show that the $\alpha^2$-dynamo becomes supercritical for $\mathcal{R}o \lesssim 10$, in reasonable agreement with the estimated threshold limit $\mathcal{R}o \lesssim 100$ inferred from observations[2] (Mohanty & Basri 2003). If this scheme is correct, the $\alpha\Omega$ (shell) dynamo might be dominant in slow rotators with strong velocity gradients at the interface between the radiative core and the convective envelope, whereas the $\alpha^2$ dynamo would dominate in fast, solid-body rotators, leading to rotation-activity saturation. This provides an appealing explanation for the observed saturated activity-rotation relationship in late M stars.

For the brown dwarf model with a conductive core and the presence of differential rotation, the onset of dynamo action is found at smaller angular velocities, i.e. longer rotational periods, for $\Omega_\star \gtrsim 0.5$, $\mathcal{R}o \lesssim 25$. The field in that case is toroidal and axisymmetric, resembling the solar field.

As shown by the present calculations, a large-scale field can be generated in fully convective objects like very late-type stars and brown dwarfs. The observed decrease of activity for later types ($\gtrsim$ M9) thus very likely stems from the lack of substantial current generation in such cool atmospheres, as explored in Mohanty et al. (2002). These authors found that the large, constant decrease in conductivity in the outermost layers of cool ($T_{eff} \lesssim 3000$ K) objects, with characteristic magnetic Reynolds numbers smaller than 1, prevents large nonpotential field configurations. Consequently, these atmospheres cannot support substantial currents ($\nabla \times \mathbf{B} = (4\pi/c)\mathbf{j} \approx 0$). The decreasing conductivity arises from both decreasing temperature and thus electron density, and ongoing grain formation and thus increasing neutral-charged particle collision frequency, knocking the charged particles off the field lines. Because of this decoupling between the magnetic field and the atmospheric fluid motions, no current is created by the field in these regions, and thus not enough shearing and twisting magnetic energy can be supplied to support a chromosphere and thus activity. Flares in late M and L dwarfs thus stem most likely from buoyant, thick flux tubes generated in the interior, where the conductivity is high enough to produce large magnetic stresses, and rising rapidly through the atmosphere, within a timescale shorter than dissipation timescales, dissipating the associated currents in the upper atmospheric layers. On the other hand, the persistent strong H$_\alpha$ emission, with H$_\alpha$ equivalent widths $> 20$ Å, observed in objects with late-M, L or even T spectra in very young associations ($\lesssim 10$ Myr) or in very few field objects ($\sim 1\%$ of the known sample) proceeds very likely from an entirely different source than from a hot chromosphere, namely external accretion (Liebert et al. 2003, Burgasser et al. 2002, Kenyon et al. 2004).

Whether the newly discovered binary brown dwarf system with quiescent X-ray emission levels of $\log L_X/L_{bol} =$ -4.2 and -4.4, respectively (Stelzer 2004), contradicts this general scheme depends crucially on the precise determination of the masses and the age of the system. Mean values for the dynamically determined masses give $M_A \simeq 0.07 \, M_\odot$, $M_B \simeq 0.055 \, M_\odot$ (Zapatero et al. 2004). Age determinations for the parent star range from 300 to 800 Myr (Stelzer 2004). A 300 Myr value places the system in the M6-M9 spectral type range, whereas a mean value of 500 Myr pushes it to the $\sim$ M8-L1 range (Chabrier & Baraffe 2000). These values fall in the $\sim$ M9-L0 region where observations show a rapid decline of the activity level (Mohanty & Basri 2003) and where the abrupt rise of resistivity is predicted (Mohanty et al. 2002). For older ages and/or smaller masses, the spectral type determination of these objects would be located in the late-L or even T domain, with effective temperatures $T_{eff} < 2000$ K, where the atmosphere is dominantly neutral and persistent activity is predicted not to occur. In that case, a possibility would be that H$_\alpha$ and X-ray activity, i.e. chromospheric and coronal emissions are not related. The persistent radio activity observed up to L3.5 (Berger 2002) may support this suggestion. The calculations of Mohanty et al. (2002), however, are based on atmosphere models of low-mass stars and brown dwarfs near the photosphere, i.e. in optical depth regimes $\tau \geq 10^{-3}$. These models do not apply to possible corona layers.

Interestingly, this general scheme for the generation and the dissipation of the magnetic field and electric currents in low-mass, fully convective objects also provides an appealing, although speculative explanation for the increasing spindown timescales with increasing spectral type, from G, K to early M and then to late M and L, i.e. with decreasing mass and effective temperature. Observations indicate that the loss of angular momentum in very-low-mass stars is smaller than that expected from a direct relation between the angular momentum loss rate $\dot{J}$ and the Rossby number (Sills et al 2000). As shown in the present calculations, the *topology of the field* generated by $\alpha^2$ dynamo differs from the organized dipole field generated by $\alpha\Omega$; the non-axisymmetry of the generated field strongly reduces the dipolar component and thus the magnetic breaking. Furthermore, the strongly decreasing conductivity in the atmosphere of cool objects hampers the formation of a hot magnetized corona and thus of magnetized winds. Both effects, weakness of the field dipolar component and increasing atmospheric resistivity, reduce the Alfvén radius $r_A$, and thus $\dot{J}$, for a given surface magnetic flux. Indeed, to a first approximation, and assuming a rigid body angular momentum, the ratio of angular momentum loss rate between a magnetized and a non-magnetized wind for a star of radius $R_*$ is $\dot{J}_{mag}/\dot{J}_{non-mag} = r_A^2/R_*^2$ for a radial field. As $r_A$ decreases and approaches a value $r_A \sim R_*$, the angular momentum loss time scale approaches the pure, non-magnetized mass loss limit for a rigid body : $\tau_J = \frac{2}{5}\frac{M}{\dot{M}}$. Assuming a maximum mass loss rate similar to the solar value, $\dot{M} \sim 10^{-14}\,{\rm M}_\odot\,{\rm yr}^{-1}$, this yields for objects at the bottom and below the main sequence ($M \lesssim 0.1\,{\rm M}_\odot$) a time scale comparable to or larger than a Hubble time, as noted previously by Giampapa et al. (1996).

---

[1] The $\alpha\Omega$ dynamo is the limit of the more general $\alpha^2\Omega$ dynamo when the magnetic Reynolds number associated with the differential rotation is much larger than that associated with the $\alpha$-effect, i.e. when this latter can be neglected as a source of a toroidal field compared with the rotational shear. So in principle, both processes are present in a star, but their relative strength depends on the strength of the differential rotation.

[2] Rossby numbers can only be *estimated* from observations and models, with large uncertainties due to the fact that (i) only the projected rotation velocity $v \sin i$ is measured, (ii) the convective timescale, pressure scale height and thus Rossby number change significantly depending at which depth they are evaluated.

An interesting possibility to verify the present theory would be Doppler imaging of fast-rotating low-mass stars and brown dwarfs. A small scale (non-helical) turbulent dynamo is likely to yield a spatially uniform chromospheric activity whereas the large-scale $\alpha^2$ process suggested in the present paper generates asymmetry, due to the absence of rotational shear. Accordingly, the present model favors a low surface filling factor for the spots rathers than a large axisymmetric polar spot to explain the recent observations of very-low-mass stars in the Pleiades (Scholz, Eislöffel & Froebrich 2005). Alternatively, a small-scale dynamo might also be present, producing evenly distributed field structures too small to be observable in the light curve. Moreover, non-axisymmetric fields can propagate in longitudinal directions without any cyclic variation of the total field energy whereas dynamo waves generated by $\alpha\Omega$ processes propagate only along the lines of constant rotation rate (Parker 1955). A consequence of the present model is that we do not expect cycles for uniformly rotating (fully convective) low-mass objects.

Future observations of activity in objects at the bottom of the main sequence and in the substellar regime, including possibly the discovered giant exoplanets, will undoubtedly constrain the general paradigm suggested in the present paper. Besides its intrinsic interest for our general understanding of dynamo action and magnetic activity, the verification of this paradigm has important implications for various fields of astrophysics, from cataclysmic variables to pre-main sequence evolution and star, brown dwarf and planet formation.

*Acknowledgements.* G.C. acknowledges the warm hospitality of the Astrophysikalisches Institute Potsdam, where this work was finalized.

## References

Barnes, J.R. et al., 2005, MNRAS, submitted
Berger, E., 2002, ApJ, 572, 503
Briggs, K.R., & Pye, J.P., 2004, MNRAS, 353, 673
Burgasser, A., Liebert, J., Kirkpatrick, D., & Gizis, J., 2002, AJ, 123, 2744
Chabrier G., & Baraffe, I., 1997, A&A, 327, 1039
Chabrier G., & Baraffe, I., 2000, ARA&A , 38, 337
Chabrier G., Baraffe, I., Allard, F., & Hauschildt, P.H., 2000, ApJ, 542, 464
Choudhuri, A.R., Schüssler, M., Dikpati, M., 1995, A&A, 303, L29
Delfosse, X., Forveille, T., Perrier, C., & Mayor, M., 1998, A&A, 331, 581
Durney, B., De Young, B., & Roxburgh, I., 1993, A&A, 331, 581
Elstner D., Meinel R., Rüdiger G., 1990, Geophys. Astrophys. Fluid Dyn. 50, 85
Ferriz-Mas, A., Schmitt, D., Schüssler, M., 1994, A&A, 289, 949
Giampapa, M., Rosner, R., Kashyap, V., Fleming, T., Schmitt, J., & Bookbinder, J., 1996, ApJ, 463, 707
Gizis, J.E., Reid, I.N., & Hawley, S.L., 2002, AJ, 123, 3356
Gizis, J.E., Monet, D., Reid, I.N., Kirkpatrick, D., Liebert, J., & Williams, R., 2000, AJ, 120, 1085
Kenyon, M.J., Jeffries, R.D., Naylor, T., Oliveira, J.M., & Maxted, P., 2004, MNRAS, in press (astro-ph/0409749)
Kitchatinov L.L., 1991, A&A, 243, 483
Kitchatinov, L.L., Rüdiger, G., 1992, A&A, 260, 494
Kitchatinov L.L., Pipin V.V., Rüdiger G., 1994, Astron. Nachr. 315, 157
Krause F., Rädler K.H., 1980, Mean-Field Magnetohydrodynamics and Dynamo Theory. Akademie-Verlag, Berlin
Krivodubskij V.N., Schultz M., 1993, Complete alpha-tensor for solar dynamo. In: Krause F., Rädler K.-H., Rüdiger G. (eds.) IAU Symp. 157, The cosmic dynamo. Kluwer, Dordrecht, p. 25
Küker, M., Rüdiger, G., 1999, A&A, 346, 922
Küker, M., Rüdiger, G., Schultz, M., 2001, A&A, 374, 301
Küker, M., Rüdiger, 2005, Astron. Nachr. , in press
Liebert, J., et al., 2003, ApJ, 125, L343
Meyer, F., & Meyer-Hofmeister, E., 1999, A&A, 341, L23
Mohanty, S., & Basri, G., 2003, ApJ583, 451
Mohanty, S., Basri, G., Shu, F., Allard, F., & Chabrier, G., 2002, ApJ, 571, 469
Moss, D., Tuominen, I., & Brandenburg, A., 1990, A&A, 240, 142
Neuhäuser, R., et al., 1999, A&A, 343, 883
Parker, E.N., 1975, ApJ, 198, 205
Parker, E.N., 1955, ApJ, 122, 293
Rädler K.-H., Wiedemann E., Brandenburg A., Meinel R., Tuominen I, A&A, 1990, 239, 413
Randich, S., 2000, *Very Low-Mass Stars and Brown Dwarfs in Stellar Clusters and Associations.*, Cambridge University Press, p. 229
Reid, I., Kirkpatrick, D., Liebert, J., Gizis, J., Dahn, C.C., & Monet, D., 2002, AJ, 124, 519
Rüdiger G., Brandenburg A., 1995, A&A, 296, 557
Rüdiger G., Kitchatinov L.L., 1993, A&A, 269, 581
Rüdiger, G., & Brandenburg, A., 1995, A&A, 296, 557
Rüdiger, G., Elstner, D., & Ossendrijver, M., 2003, A&A, 406, 15
Rüdiger, G., & Hollerbach, R., 2004, *The magnetic Universe*, Wiley
Scholz, A., Eislöffel, J., & Froebrich, D., 2005, A&A, 438, 675
Sills, A., Pinsonneault, M., & Terndrup, D., ApJ 534, 335
Spiegel, E., & Weiss, N., 1980, Nature, 287, 616,
Spiegel, E., & Zahn, J.-P., 1992, A&A, 265, 106
Stauffer, J., Giampapa, M., Macintosh, B., Reid, I., & Hamilton, D., 1994, AJ, 108, 160
Stelzer, B, 2004, ApJ, 615, L153
Stix, M., 1976, A&A, 47, 243
Tsuboi, Y., Maeda, Y., Feigelson, E., Garmire, G., Chartas, G., Mori, K., & Pravdo, S., 2003, ApJ, 587, L51